\documentclass[superscriptaddress,amsmath,amssymb, aps, pra]{revtex4-2}

\usepackage{graphicx}
\usepackage{dcolumn}
\usepackage{bm}
\usepackage[colorlinks]{hyperref}
\usepackage[T1]{fontenc}
\usepackage[utf8]{inputenc}
\usepackage{dblfloatfix}
\usepackage{amsmath}
\usepackage{float}
\usepackage{braket}
\usepackage{url}
\usepackage{amsfonts}
\usepackage{bbold}
\usepackage{graphicx}
\usepackage{caption}
\usepackage{subcaption}
\usepackage{appendix}
\usepackage{comment}
\usepackage{lastpage}
\usepackage{float}
\usepackage{tabularx}
\graphicspath{ {images/} }
\usepackage{fancyhdr}
\usepackage{graphicx}
\usepackage{caption}
\usepackage{subcaption}
\usepackage{lastpage}
\usepackage{qcircuit}
\usepackage{titlesec}
\titlespacing*{\section}{0pt}{1.2\baselineskip}{\baselineskip}

\begin{document}

\title{Variational Quantum Algorithm based circuit that implements the Toffoli gate with multi inputs}

\author{Yuval Idan}
\affiliation{The Alexander Kofkin Faculty of Engineering, Bar-Ilan University, Ramat Gan 5290002, Israel}

\author{M.N.Jayakody}
\affiliation{The Alexander Kofkin Faculty of Engineering, Bar-Ilan University, Ramat Gan 5290002, Israel}

\begin{abstract}
The prime objective of this study is to seek a circuit diagram for a multi-inputs Toffoli gate including only single qubit gates and CNOTs. In this regard, we have developed two variational quantum algorithms (VQAs) that can be used to implement a multi-inputs Toffoli gate. The cost functions of these two VQAs are derived by using the Hilbert–Schmidt inner product and the expected value of an observable that can capture the difference between the inputs and outputs of a Toffoli gate. We employ two ansatz circuit architectures and use the PennyLane package to execute the optimization. 
\end{abstract}

\maketitle

\section{Introduction}


Variational Quantum Algorithms (VQAs) have become one of the key tools that enable us to harness the quantum advantage from Noisy Intermediate Scale Quantum (NISQ) devices. VQAs can be viewed as the quantum analog of classical machine-learning methods. The core concept of the VQAs is the optimization of a parameterised quantum circuit, which is designed to be run on a quantum computer yielding a desired output. Moreover, VQAs adopt a leading strategy of outsourcing the parameter optimization to a classical optimizer in order to by pass the constraints imposed by NISQ devices that diminish the quantum advantage. As a result, in VQAs the depth of the quantum circuit can be reduced substantially and the noise mitigation becomes easier compared to that of quantum algorithms developed for the fault-tolerant era \cite{cerezo2021variational}. Variational Quantum Eigensolver (VQE) \cite{peruzzo2014variational} and Quantum Approximate Optimization Algorithm (QAOA) \cite{farhi2014quantum} are the best known applications of the VQAs. The prior estimates the ground-state molecular energy for $He-H^{+}$ by combining a highly reconfigurable photonic quantum processor with a conventional computer and the latter approximately solves combinatorial problems such as Constrain-Satisfaction and Max-Cut problem. 

In this project we intend to make use of a variational quantum algorithm to implement a Toffoli gate with $n$ input qubits that constitute a single target and $n-1$ control qubits. Such a Toffoli gate can be termed as a multi-controlled Toffoli (MCX) gate. In its internal structure, a MCX must include many quantum operations in order to yield the desired output. It can be stated with certainty that when the number of control qubits increases, the complexity of the internal structure of the MCX also increases proportionally. Hence, it is not straightforward to decompose a MCX gate with $n$ input lines into units of single qubit gates and CNOT gates. To resolve such a problem, VQAs can be considered as a potential candidate. Therefore, we intend to develop a reliable VQA with the aim of determining the optimum internal gate structure for a MCX with $n$ inputs under the constrain that a single unit of the internal structure comprises either a one qubit gate or a CNOT gate. We develop two separate VQAs and test them for Toffoli gates with 3 and 5 inputs.  

\section{Basic concept and the components of VQAs}

\begin{figure*}[ht!]
\centerline{\includegraphics[width=4in, height=2in]{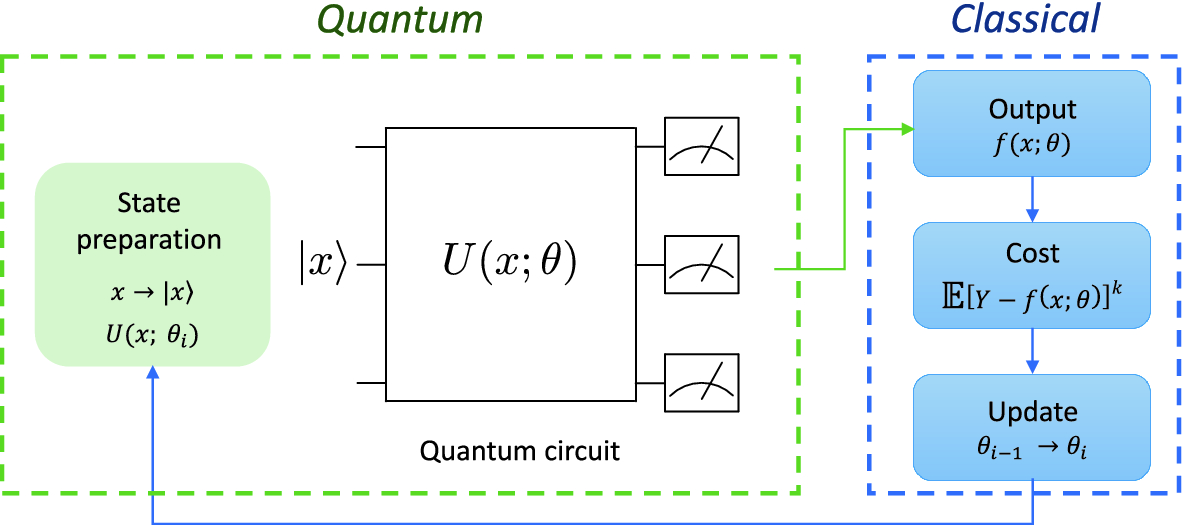}}
\caption {Schematic diagram of a VQA. The initial state $|x\rangle$ is prepared and fed as an input to the parameterized ansatz which is characterized by the unitary operator $U(x,\theta)$. Next, the measured outcome is inserted to a classical optimizer in order to determine the optimal values for the set of parameters of $\theta$'s in such a way that the cost function $f(x,\theta)$ becomes minimum as possible. Finally, the parameters of $\theta$s are updated and the process is repeated until an optimal solution is attained. Source \cite{macaluso2020variational} }
\label{fig_1}
\end{figure*}

The variational method in quantum theory paves a way to find the lowest energy state or the ground state of a given quantum system  \cite{sommerfeld2011lorentz}. In this method, a trial wave function that comprises one or more parameters is chosen at first as an approximation to the ground state of the system. Sometimes this trial wave function is termed as "ansatz". Then, using the variational principle, the optimal values for the parameters of the trial wave function are obtained in such a way that the expectation value of the energy becomes lowest as possible \cite{griffiths2018introduction}. The aforementioned variational method lies at the heart of the VQAs. One can identify three main components that constitute a VQA. They are, the cost function, ansatzes and the optimization. Schematic diagram of a variational quantum algorithm that includes theses main three components is given in Figure \ref{fig_1}. 
\subsection{Cost function}
In the context of classical machine leaning, the cost function can be understood as the technique of evaluating how well the machine learning model performs for a given dataset \cite{raschka2019python}. In general, the cost function determines the difference between the expected and predicted values of the machine learning model and quantifies it in terms of a single real number. Similarly, in a VQA, encoding the problem of interest into a cost function is one of the important phases of its development. For a VQA, the cost function can be expressed in a more general way as follows
\begin{equation}\label{general_cost_fucntion}
    C(\boldsymbol{\theta})=f\left(\{\rho_k\},\{\mathcal{O}_k\},U(\boldsymbol{\theta})\right)
\end{equation}
where $f$ is some function, $U(\boldsymbol{\theta})$ is a parameterized unitary, $\boldsymbol{\theta}$ is composed of discrete/continuous parameters, $\{\rho_k\}$ are input states from a training set and $\{\mathcal{O}_k\}$ are a set of observable \cite{cerezo2021variational}. For our study, we need to choose a suitable cost function that can compare the difference between the input and output states of the $n$-inputs Toffoli gate. Since the matrix form of the $n$-inputs Toffoli gate is well known, we employ the Hilbert Schmidt test to derive a suitable cost function to cater the demand of our problem. In addition, as a second method, we derive an observable combining a set of projectors that can capture the difference between the input and output states of a multi-input Toffloi gate. Then, by calculating the expected value of this special operator we determine the cost function. 
\subsection{Ansatz}
Ansatz is a well-known term in mathematics that describe an educated guess made to help in solving a given problem. Since we work with quantum machine learning methods, it is crucial to define our ansatz in a particular way that is unknown at the beginning of our project. We test our code on two types of ansatz configurations named basic entangled layer and strongly entangled layer and determine what is the most suitable configurations for our study. The stated goal of our project is to determine the internal gate structure of a $n$-inputs Tofolli gate that comprises only single qubit gates and CNOTs. Hence, when adopting the aforementioned two different ansatz configurations, we use the most general single qubit gate of the following form to represent any single qubit gates
\begin{equation}\label{coin_matrix}
U(\theta,\phi,\lambda)=\left(
\begin{array}{cc}
 \cos{\theta/2} & -e^{i\lambda}\sin{\theta/2}  \\
 e^{i\phi}\sin{\theta/2} & e^{i(\theta +\phi)}\cos{\theta/2}  \\
\end{array}
\right)
 \end{equation}
where $\theta \in [0,2\pi)$ and $\phi,\lambda \in [0,\pi]$. In qiskit and Pennylane the most general single qubit gate is given by $U3$ operator.
\subsection{Gradients and Optimizers}
Optimization process of a VQA makes use of the gradients and optimizers to speed up and to guarantee the convergence. For many optimization tasks, the information about the gradient of the cost function can be used to speed up the convergence of the optimization \cite{cerezo2021variational}. In this regard, the method of parameter-shifting rule is used to calculate the first and higher-order derivatives of the cost function. Optimizers are objects which can be used to automatically update the parameters of a quantum or hybrid machine learning model. For our study, we use \textit{GradientDescentOptimizer}, one of the gradient-based optimizers available in the PennyLane package. All the gradient-based optimizers provided by PennyLane update the variational parameters moving along the direction indicated by the gradient at the specified point. \textit{GradientDescentOptimizer} is the simplest version of the gradient-based optimizers available in PennyLane. 
\section{Toffoli gate}
\begin{figure}[ht!]
    \includegraphics[width=6in, height=1.5in]{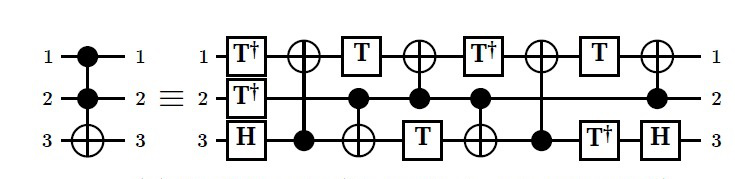}
    \caption{The circuit diagram of the Toffoli gate with 3 inputs according to \cite{nielsen2002quantum}. This implementation is proved to be the most efficient decomposition}
    \label{3-input_Toffoli gate}
\end{figure}

The CNOT gate operates on a quantum register consisting of 2 qubits. The CNOT gate flips the second qubit (the target qubit) if and only if the first qubit (the control qubit) is ${\displaystyle |1\rangle }|1\rangle$. An expansion to the CNOT gate is the ``Toffoli gate".
Toffoli gate is an ``AND'' gate that contains $N-1$ control qubits and one target qubit. For an arbitrary quantum state $\ket{\psi}=\bigotimes_{i=1}^{N}\ket{i}$, using the Toffoli gate $\hat{\mathcal{T}}$ the state will be transformed to the following state
\begin{equation}
\hat{\mathcal{T}}\ket{\psi}=\hat{\mathcal{T}}\bigotimes_{i=1}^{N}\ket{i}\rightarrow  \bigotimes_{i=1}^{N-1}\ket{i}\otimes\ket{\prod_{k=1}^{N-1}k+N \ mod_2} 
\end{equation}
Thus, we will add one modulo 2(in dimensional 2 of Hilbert's space, we can work in higher Hilbert's spaces, which means that we will work with different groups ) to the target qubits if and only if all of  the control qubits are 1. We can implement Toffoli gate with several configurations. However, in our project we want to do so only with single qubit gates and CNOT gates (see Figure \ref{3-input_Toffoli gate})(it is possible to do so only with CNOT but it is not efficient in terms of quantum depth). Our goals is to create multi inputs Toffoli gate with quantum machines learning methods and find a way to minimize the circuits depth. The depth of a quantum circuits is the amount of operators in the longest path in the circuits.
\subsection{Implementation of Toffoli gate}
The basic Toffoli gate (3 inputs) can be arrange according to \cite{nielsen2002quantum} with depth of 10. When the Toffoli gate contains more inputs, the number of needed gates does not increase linearly and is very hard to create (this is why Classisq announced the Toffoli's challenge). On the other hand, we noticed that the Toffoli gate contains two parts; the first part is the ``AND" gate, and the second part is the phase fix (we want to affect only the target qubit). With the above prior knowledge, we can construct our layers in a certain way and improve the running time of our algorithm.
e.g., We can for the first part, we can use a linear entangled pattern, and for the second part pattern we can use a maximally entangled pattern.
\cite{toffoli1980reversible}
\section{First Method: Hilbert Schmidt test}
The first step in our implementation is to try to copy the ``generic''  Toffoli gate (using an existing function that is not necessarily composed of just single qubit gates and CNOTS) with VQE using the Hilbert Schmidt test (HST) as a cost function.
We succeeded in this task for small-scale operations (2-5 qubits).
The HST was introduced in  \cite{khatri2019quantum}, it takes two operators and calculates a metric of their difference.
To be more precise, it calculates the following quantity:
\begin{equation}\label{Hilbert Schmidt test}
    C_{HST}=1-\frac{1}{d^2}|\braket{V,U}|^2=1-\frac{1}{d^2}|Tr(V^{\dagger}U)|^2
\end{equation}
where $d$ is the dimension of Hilbert space.
In other words, by using the HST as the cost function of our learning process we're able to acquire an approximation of the target circuit.
\begin{figure}[ht!]
    \centering
    \includegraphics[width=4in, height=2in]{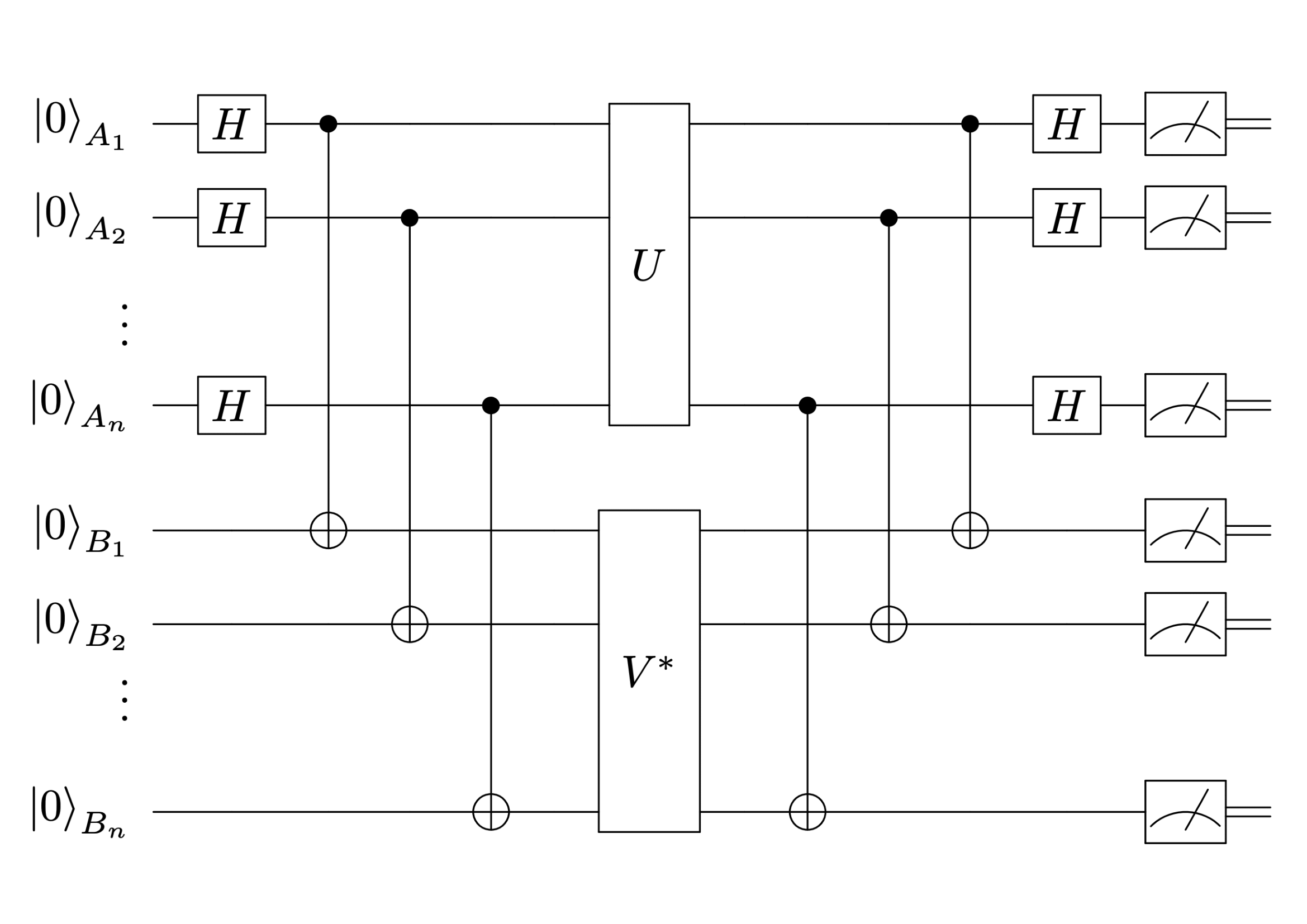}
    \caption{The schematic circuit diagram for the Hilbert Schmidt test. Unitary $U$ is the Toffoli function and $V$ is our antaz circuit. For each qubit in the Toffoli gate we need to implement two qubits one for $U$ and  one for $V$.}
    \label{Hilbert Schmidt test}
\end{figure}
Adding rotations to each layer, each layer is composed of 3 operators.
We used four layers but removed two of the cnot layers at the end. Thus we created a circuit with a depth of 10.
We simulated the same optimizations with random initial parameters and succeeded in imitating the Toffoli gate according to the Hilbert Schmidt test.
We should note that our circuit has a greater depth than the original Toffoli gates, but we can imitate unknown circuits with only rotations and CNOT.
\begin{figure}[h]
  \centering
     \begin{subfigure}[b]{0.458\textwidth}
         \centering
         \includegraphics[width=\textwidth]{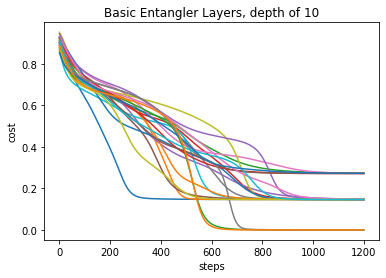}
         \caption{Results for Toffoli gate with three qubits using Hilbert Schmidt test as a cost function, The layering method is ``Basic entangled layering''}
         \label{fig:3 qubits}
     \end{subfigure}
     \hfill
     \begin{subfigure}[b]{0.458\textwidth}
         \centering
         \includegraphics[width=\textwidth]{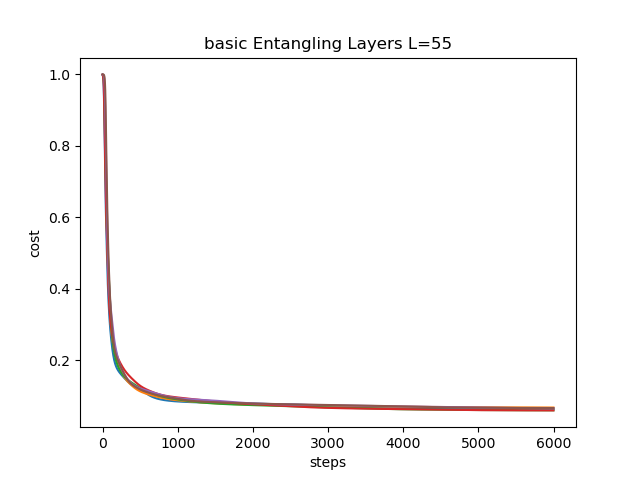}
         \caption{Results for Toffoli gate with five qubits using Hilbert Schmidt test as a cost function, The layering method is ``Basic entangled layering''}
         \label{fig:5 qubits}
     \end{subfigure}
     \caption{Results for the Toffoli gates with 3 and 5 input qubits using the Hilbert Schmidt test as the cost function }
     \label{figure_resuits_of_Hilbert_test}
\end{figure}
\subsection{Testing our circuits}
\begin{figure}
    \centering
    \includegraphics[width=4in, height=3in]{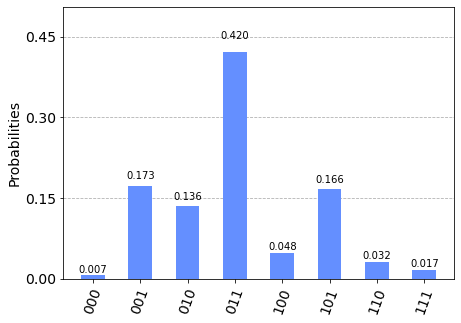}
    \caption{Simulation's results for our cloned Toffoli gate with three qubits, for \(\epsilon\leq 0.000001\), we can see that for the state \(\ket{111}\) we are getting: \(\ket{\psi}=\sqrt{0.58}\ket{else}+\sqrt{0.42}\ket{110}\)}, the state with the higher probability is the correct state after the Toffoli gate, but most of the time we will get incorrect state.
    \label{Simulating our results} 
\end{figure}

The learning algorithm stops when the condition \(\epsilon\leq 0.000001\) is met. By choosing a smaller value for $\epsilon$ we could obtain a more accurate approximation but it would also increase the time it takes for the algorithm to complete.
To check if our method is working we saved our weights from the HST and tested it on ``qiskit'' code and check all the state to be sure how good is our approximate.
The next step was to transform our Pennylane results to a qiskit circuit and using a simulator in order to test the results of the circuits.
Unfortunately, the initial \(\epsilon\) is not good enough and not usable in real quantum circuits (Figure \ref{Simulating our results}).  
From the results we can see that the state with the biggest amplitude is the correct state after the Toffoli gate ,but unfortunately it's not good enough and can`t be implemented in a real quantum circuit. We can take smaller \(\epsilon\) but it will take more layers which would increase the depth of our circuits.
We can conclude up that the Hilbert Schmidt test may be not the best method to create Toffoli gate because for insufficently small values of $\epsilon$ the approximation is not good.
We will preform a second method in order to try and get better results.
\subsection{Toffoli gate with 5 qubits}
When approaching the limit of seven qubits we noticed and dealt with an exponential jump in the run time and the number of iterations and layers.
We successfully managed to imitate Toffoli gates of 5 and 7 qubits perfectly with fidelity of 80 percent.
We should note that imitating a Toffoli gate with more than 7 qubtis isn't possible for our computers at a reasonable ammount of time (We used intel 12th Gen Intel(R) Core(TM) i7-12700KF 3.61 GHz).
In addition, the number of layers that we need in order to imitate the Toffoli gate became inefficient as we can see in Figure \ref{fig:5 qubits}, we couldn't reach the global minima in reasonable running time.
\section{Second method: Defining cost function in terms of the expected value}
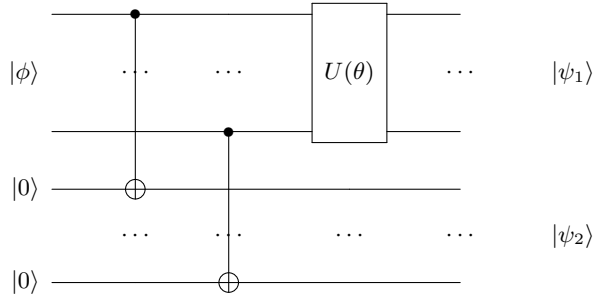
\begin{figure}
    \[ \Qcircuit @C=3em @R=1.5em { 
\lstick{} & \ctrl{3} & \qw & \multigate{2}{U(\theta)} & \qw  \\ 
\lstick{\ket{\phi}} & \cdots & \cdots & \nghost{U(\theta)} & \cdots  & & \lstick{\ket{\psi_1}} \\
\lstick{} &\qw & \ctrl{3} & \ghost{U(\theta)} & \qw \\
\lstick{\ket{0}} & \targ  &\qw & \qw & \qw  \\
\lstick{} & \cdots & \cdots &  \cdots & \cdots  & & \lstick{\ket{\psi_2}} \\
\lstick{\ket{0}}  &\qw  & \targ & \qw & \qw  
} \]
\caption{The schematic circuit diagram for the second method}
    \label{Schematic_circuit_diagram_second_method}
\end{figure}

\begin{figure}
    \centering
    \includegraphics[width=6in, height=2in]{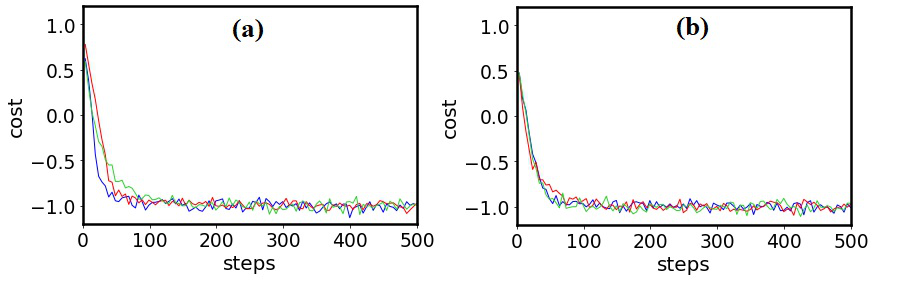}
    \caption{The minimization of the cost function in the second method for (a) basic entangled (b) strongly entangled architectures with 8 layers for the case of 3-inputs ansatz circuit. Optimization is done of 500 steps. Different colours represent graphs for three different initial states.} 
    \label{ansatz_for_3-input_Toffoli gate} 
\end{figure}

In the previous approach, the cost function is derived from the Hilbert–Schmidt inner product as shown in  \eqref{Hilbert Schmidt test}. For this calculation, we take the trace of the product of parameterized unitary operator that represents the ansatz and the matrix form of the multi-input Toffoli gate. As an alternative method to the Hilbert–Schmidt test, in this section, we intend to derive another type of cost function in terms of the expected value of an observable that can capture the difference between input and output states of the ansatz circuit. Then, by minimizing the expected value of this observable we expect to obtain the desired optimization in the ansatz circuit. The schematic circuit diagram for the second method is given in Figure \ref{Schematic_circuit_diagram_second_method}. According to the circuit diagram given in Figure \ref{Schematic_circuit_diagram_second_method}, the first register is linked to the circuit ansatz represented by the parameterized unitary operator $U(\theta)$. The input and the output states of the ansatz are given by $|\phi\rangle$ and $|\psi_1\rangle$ respectively. The second register is initiated from the $|\mathbf{0}\rangle$ and the final state of the second register is represented by $|\psi_2\rangle$. Then, the state $|\psi_f\rangle$ of the whole circuit can be written as
\begin{equation}\label{fianl_state}
    |\psi_f\rangle=|\psi_2\rangle \otimes |\psi_1\rangle
\end{equation}
Note that, initially, the first and second registers get entangled due to the configuration of CNOT gates. Thus $|\psi_2\rangle = |\phi\rangle$. That is $|\psi_2\rangle$ and $|\psi_1\rangle$ give the input and the output states of the ansatz circuit respectively. Therefore, by comparing $|\psi_2\rangle$ and $|\psi_1\rangle$ we can identify whether the ansatz circuit gives the correct output state for the corresponding input state. \\
Our next task is to define a suitable operator in such a way that the expected value of this operator can be used to make the distinction between $|\psi_1\rangle$ and $|\psi_2\rangle$ efficiently. Let us define a set $R_n$ including the pairs of input $|\chi_{in}^{(k)}\rangle$ and the corresponding output $|\chi_{out}^{(k)}\rangle$ states of a Toffoli gate with $n$ input qubits as $R_n=\bigg\{ \left(|\chi_{in}^{(k)}\rangle ,|\chi_{out}^{(k)}\rangle\right)\bigg\}_{k=0}^{2^n-1}$. Now define an operator of the following form
\begin{equation}\label{Operator}
    A_n=\mathbb{I}-2\sum_{k=0}^{2^n-1}{|\chi_{in}^{(k)}\rangle\langle\chi_{in}^{(k)}| \otimes |\chi_{out}^{(k)}\rangle\langle\chi_{out}^{(k)}| }
\end{equation}
Note that, since $A_n=A^{\dagger}_n$, the operator given in \eqref{Operator} is Hermitian. Thus, we can consider $A_n$ as an observable. The expected value of $A_n$ can be written as

\begin{equation}\label{Expected_val_Operator}
\langle \psi_f|A_n|\psi_f \rangle=
   \begin{cases}
      1 & \left(|\psi_2\rangle ,|\psi_1\rangle\right)  \not\in R_n \\
     -1 & \left(|\psi_2\rangle ,|\psi_1\rangle\right) \in R_n\\
   \end{cases}
\end{equation}
The proof of the expected value in \eqref{Expected_val_Operator} is given below. Write $G=\sum_{k=0}^{2^n-1}{|\chi_{in}^{(k)}\rangle\langle\chi_{in}^{(k)}| \otimes |\chi_{out}^{(k)}\rangle\langle\chi_{out}^{(k)}|}$. Then we have $A_n=\mathbb{I}-2G$. Suppose the circuit ansatz in Figure \ref{Schematic_circuit_diagram_second_method} yields a correct input and output pair. Then, for some unique $r \in [0, \hdots, 2^n-1]$, we have $|\psi_1\rangle=|\chi_{out}^{(r)}\rangle$ and $|\psi_2\rangle=|\chi_{in}^{(r)}\rangle$. Hence, the expected value of operator $G$ can be written as $\langle \psi_f|G|\psi_f \rangle=\left(\langle \chi_{in}^{(r)}|\otimes\langle \chi_{out}^{(r)}|\right)G\left(|\chi_{in}^{(r)} \rangle \otimes|\chi_{out}^{(r)} \rangle\right)=1$. Thus we have  $\langle \psi_f|A_n|\psi_f \rangle=-1$. Now suppose that the input and output pair $\left(|\psi_2\rangle ,|\psi_1\rangle\right)$ of the ansatz is not in $R_n$. Then it is obvious that $\langle \psi_f|G|\psi_f \rangle=0$. Thus we have  $\langle \psi_f|A_n|\psi_f \rangle=1$. This complete the proof. Let us now employ the second method to develop a VQA with the intention of implementing a Toffoli gate with 3 input qubits and 5 input qubits. For this, we need to make a slight modification to the schematic circuit diagram given in Figure \ref{Schematic_circuit_diagram_second_method} in order to generate all possible input states. We initiate the first register with $|\bf{0}\rangle$ and then apply Hadamard gates to each wire of the first register. Then, the state $|\phi\rangle$ represents the superposition of all possible input states of a Toffoli gate with 3 or 5 inputs. We use the Pennylean package to develop the VQAs for 3 and 5 input Toffoli gates. Derivation of the observable $A_3$ and $A_5$ is given in Appendix \ref{Derivation_of_A3_A5}. Moreover, we consider two parameterized ansatz circuit layers. In the first case we build the circuit ansatz using the basic entangle layer architecture given in Figure \ref{basic_entangled_layer}. Next, the strongly entangled layers architecture given in Figure \ref{stronly_entangled_layer} is utilized. The graphs related to the minimization of the cost function in both architectures are given in Figure \ref{ansatz_for_3-input_Toffoli gate}. 
According to the graphs given in Figure \ref{ansatz_for_3-input_Toffoli gate}, after the $100^{th}$ steps, the value of the cost function tends to oscillate around $(-1)$. Hence, in order to test the efficiency of the second method, we stopped the optimization process whenever the cost function hits $(-1)$ and then saved the set of optimized parameters corresponding to the $(-1)$ of the cost function. Since the case of 5-inputs anzatz circuit is time consuming, we have followed this testing procedure only for the 3-inputs ansatz circuits. Moreover, we have considered only the basic entangled architecture with 8 layers. After picking up the set of optimized parameters corresponding to $(-1)$ of the cost function, we have run the 3-inputs ansatz circuits in IBM quantum platform by inserting this set of optimized parameters. Then, for all possible inputs, we have compared the outputs of the 3-inputs ansatz circuits with that of a 3-inputs Toffoli gate. Our optimized circuit gives the results of 3-inputs Toffoli gates for all the possible input states with more than $97\%$ accuracy. Figure \ref{optimzied_3-input_Toffoli gate} shows the optimized circuit diagrams with 8 layers that mimics the 3-inputs Toffoli gates. Some of the output results of this optimized circuit is given in Figure \ref{results_of_the_optimized_4layer_of_3input Toffloi}.
\section{Conclusion}
In this study we have developed two separate VQAs that can be used to determine a quantum circuit for a multi-input Tofolli gate consisting of single qubit gates and CNOTS. The main distinction between these two VQAs arises from the derivation of their cost functions. The VQAs of the first and second methods use the Hilbert–Schmidt inner product and the expected value of an observable to derive efficient cost functions respectively. From the perspective of run time efficiency, the first method surpasses the second method substantially. However, when it comes to the accuracy level of the optimized circuit, second method yields more favourable results. By analysing the convergence rate of the graphs given in Figures \ref{figure_resuits_of_Hilbert_test} and \ref{fig:my_label2}, it can be stated that the basic entangled layer is the most suitable architecture for fast convergences than the strongly entangled layer. Due to the run time constrains imposed by the PennyLane package, it is difficulty to run the ansatz circuits with higher number of input qubits. However, we anticipate that the outcome of this study may pave a way to develop more efficient VQAs for implementing Toffoli gates with more inputs. We experience the computational power of quantum computer by trying to implement more then 3 qubit circuit(The run time become unbearable) and  hope that in the future we could use a real quantum computer to test our results.
\bibliography{citations}
\appendix
\section{Layers}
We compared several methods of layering our ansatz. As we mentioned before, the first method is the ``basic entangled layers'' along with `` strongly entangled layers'' and  ``random layers''. The ``basic entangled layers'' are composed of a repeating pattern of unitary rotation (U3) and CNOTS that operate with neighboring(the first with the second and go on..  and the last CNOT act on the first qubit). This configuration is similar to the implementation of the Toffoli gate in that each qubit is entangled with the other and resembles the repetition of the Toffoli(Fig 2). This pattern had the best performance. We can assume that it is because of the similarities with the three qubits Toffoli gate(The architecture  look vert similar). The `` strongly entangled layers'' is similar to the ``basic entangled layers'' but in each layer the CNOT act with other pair of qubits which entaneled directly each qubit with the another. We simulate method one with strongly entangled layers and got the following results:
\begin{figure}[ht!]
    \centering
    \includegraphics[width=5.5in, height=2in]{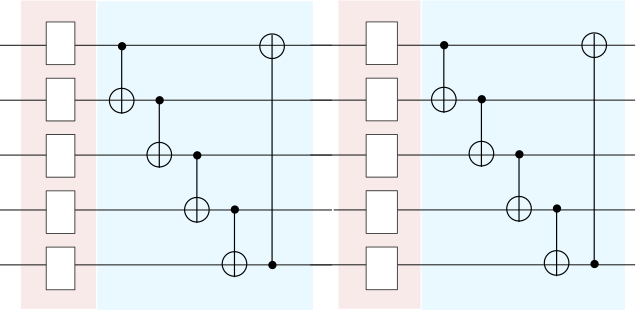}
    \caption{Basic entangled layers architecture , the best layering for our problem}
    \label{basic_entangled_layer}
\end{figure}

\begin{figure}[ht!]
    \centering
    \includegraphics[width=6in, height=2in]{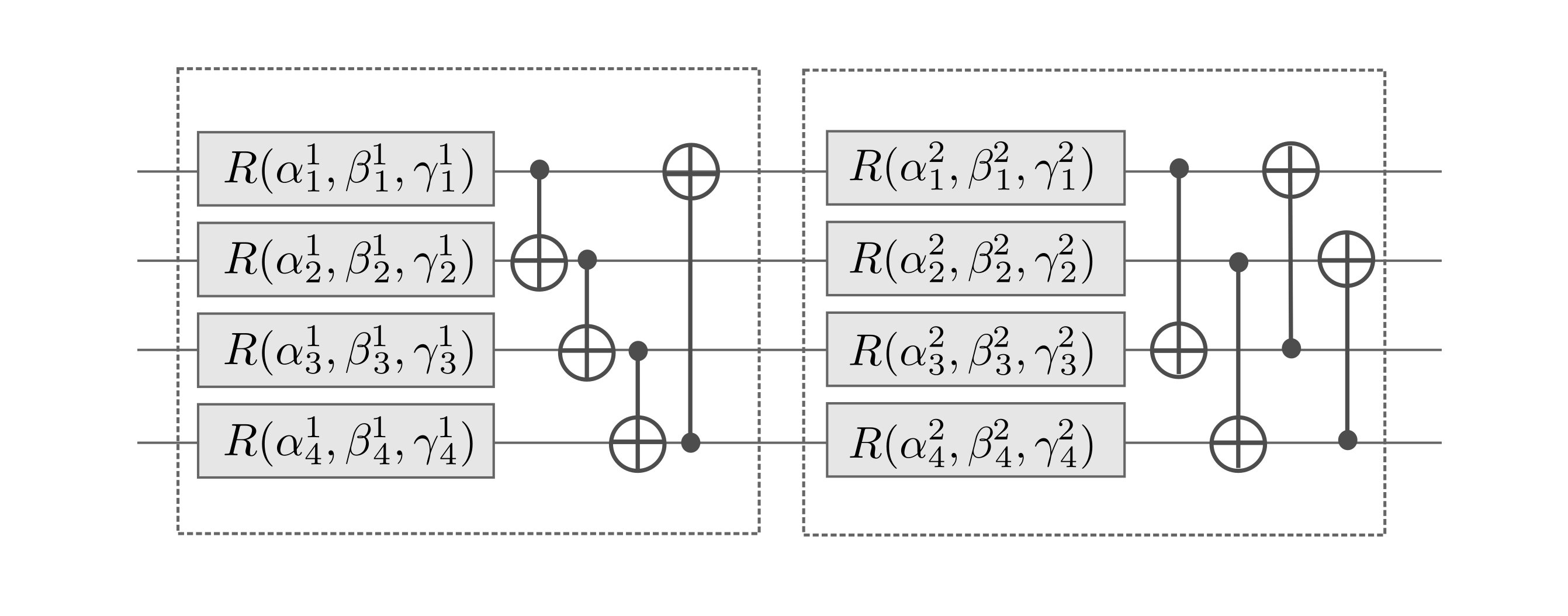}
     \caption{Strongly entangled layers architecture}
    \label{stronly_entangled_layer}
\end{figure}
\begin{figure}[ht!]
    \centering
    \includegraphics[width=6in, height=3in]{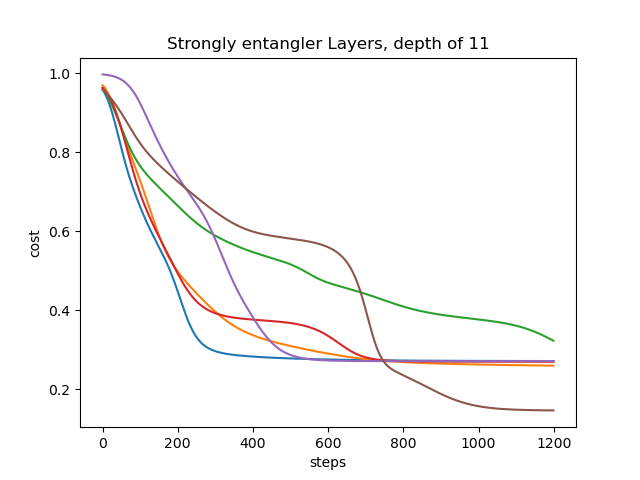}
    \caption{Strongly entangled layers results, we can see that it is less effective than the basic entangled layers}
    \label{fig:my_label2}
\end{figure}

As we expected, the results of the `` strongly entangled layers'' were less effective than the ``basic entangled layers'' as we can see in the Figure \ref{fig:my_label2}. 
\section{Derivation of the observable $A_3$ and $A_5$} \label{Derivation_of_A3_A5}
In this section, we present the detailed procedure followed to implement the Toffoli gates with 3 and 5 inputs. The $R_3$ set that includes all the pairs of inputs and their corresponding outputs of a Toffoli gate with 3 qubits can be given as 
\begin{equation}\label{R3 set}
\begin{split}
        R_3&=\{(|000\rangle,|000\rangle), (|001\rangle,|001\rangle),(|010\rangle,|010\rangle),(|011\rangle,|011\rangle), \\ &(|100\rangle,|100\rangle),(|101\rangle,|101\rangle),(|110\rangle,|111\rangle),(|111\rangle,|110\rangle)\}
\end{split}
\end{equation}
Now, let us write the explicit form of the observable $A_3$ that can capture the difference between the input and output states of a Toffoli gate with 3 qubits as  
\begin{equation}\label{A_3 operator}
\begin{split}
        A_3&=\mathbb{I}-2\bigg(\textcolor{red}{|000\rangle \langle 000|}\otimes |000\rangle \langle 000|+\textcolor{red}{|001\rangle \langle 001|}\otimes |001\rangle \langle 001|+\textcolor{red}{|010\rangle \langle 010|}\otimes |010\rangle \langle 010| \\
        &+\textcolor{red}{|011\rangle \langle 011|}\otimes |011\rangle \langle 011|+\textcolor{red}{|100\rangle \langle 100|}\otimes |100\rangle \langle 100|+\textcolor{red}{|101\rangle \langle 101|}\otimes |101\rangle \langle 101| \\
        &+\textcolor{red}{|110\rangle \langle 110|}\otimes |111\rangle \langle 111|+\textcolor{red}{|111\rangle \langle 111|}\otimes |110\rangle \langle 110|\bigg)
\end{split}
\end{equation}
Note that, the outter products marked in red and black colours correspond to input and output states respectively. From the properties of tensor products, we can rewrite the $A_3$ operator in \eqref{A_3 operator} as follows
\begin{equation}\label{A_3 operator1}
\begin{split}
        A_3&=\mathbb{I}-2\bigg(|\textcolor{red}{000}000\rangle \langle \textcolor{red}{000}000|+|\textcolor{red}{001}001\rangle \langle \textcolor{red}{001}001|+|\textcolor{red}{010}010\rangle \langle \textcolor{red}{010}010|+|\textcolor{red}{011}011\rangle \langle \textcolor{red}{011}011| \\
        &+|\textcolor{red}{100}100\rangle \langle \textcolor{red}{100}100|+|\textcolor{red}{101}101\rangle \langle \textcolor{red}{101}101|+|\textcolor{red}{110}111\rangle \langle \textcolor{red}{110}111|+|\textcolor{red}{111}110\rangle \langle \textcolor{red}{111}110|\bigg)
\end{split}
\end{equation}
When building-up $A_3$ observable in PennyLean, we use the \textit{qml.Projector} function. Consider an arbitrary state of $|abc\rangle$ where $a,b,c$ are binary variables which can take either 0 or 1. The function \textit{qml.Projector} allows us to build the projector of $|abc\rangle \langle abc|$. Hence, by deriving a customized Hamiltonian using the projectors in \eqref{A_3 operator1}, we can build-up the $A_3$ operator. Similarly, for the case of Toffoli gates with 5 inputs, the $R_5$ set can be written as follows
\begin{equation}\label{R5 set}
        R_5= \bigg\{(|i\rangle,|i\rangle)\bigg\}_{i=0}^{2^5-3} \cup \bigg\{(|30\rangle,|31\rangle), (|31\rangle,|30\rangle)\bigg\}
\end{equation}
Note that, for the sake of convenience, we have employed the decimal representation to write the input output pairs. The explicit form of the observable $A_5$ that can capture the difference between the input and output states of a Toffoli gate with 5 qubits can be written as  
\begin{equation}\label{A_5 operator}
        A_5=\mathbb{I}-2\bigg(\sum_{i=0}^{2^5-3}\textcolor{red}{|i\rangle \langle i|}\otimes |i\rangle \langle i|+\textcolor{red}{|30\rangle \langle 30|}\otimes |31\rangle \langle 31|+\textcolor{red}{|31\rangle \langle 31|}\otimes |30\rangle \langle 30|\bigg)
\end{equation}
Following the same fashion described above, we can derive a customized Hamiltonian for the operator in \eqref{A_5 operator} using PennyLean package and then calculate the expected value of it.
\begin{figure}
    \centering
    \includegraphics[width=7in, height=3in]{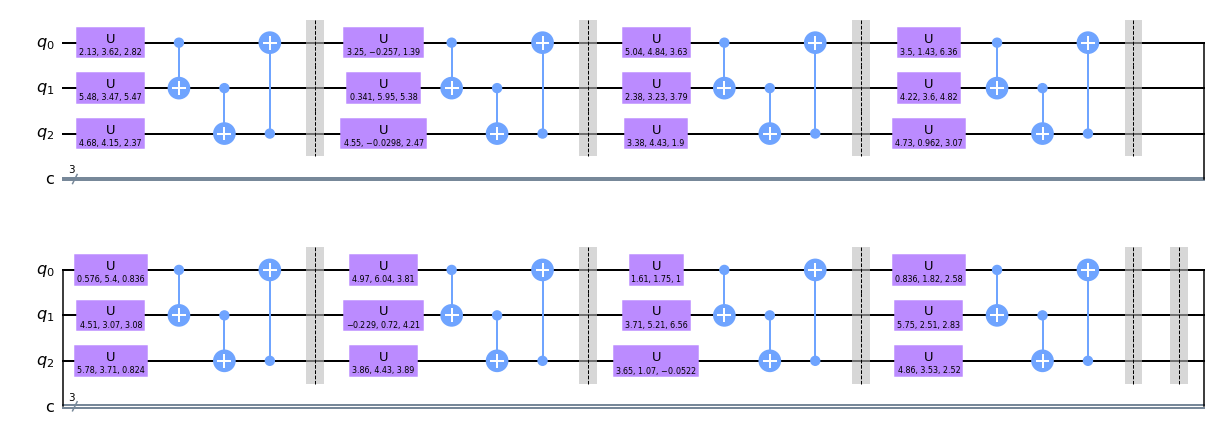}
    \caption{The circuit diagram of the optimized 3-inputs ansatz circuit in basic entangled architecture (with 8 layers) that yields the outputs of a 3-inputs Toffoli gate with $97\%$ accuracy. The optimization is done using second method.} 
    \label{optimzied_3-input_Toffoli gate} 
\end{figure}

\begin{figure}
  \centering
     \begin{subfigure}[b]{0.458\textwidth}
         \centering
         \includegraphics[width=\textwidth]{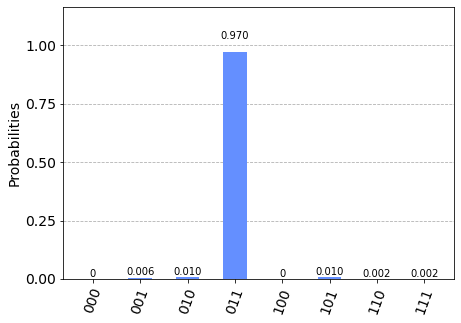}
         \caption{Output of the optimized ansatz for the input of $|111\rangle$}
     \end{subfigure}
     \hfill
     \begin{subfigure}[b]{0.458\textwidth}
         \centering
         \includegraphics[width=\textwidth]{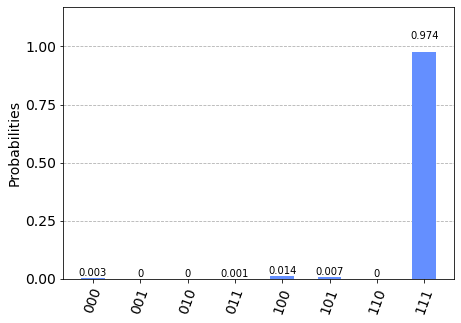}
         \caption{Output of the optimized ansatz for the input of $|110\rangle$}
     \end{subfigure}
     \caption{Some output results of the circuit diagram in Figure \ref{optimzied_3-input_Toffoli gate} (optimized 3-input ansatz in basic entangled architecture with 8 layers). The optimization is done using second method.}
     \label{results_of_the_optimized_4layer_of_3input Toffloi}
\end{figure}

\end{document}